\documentclass[journal,comsoc,10pt]{IEEEtran}

\usepackage[T1]{fontenc}

\usepackage{makecell}
\usepackage{cite}
\usepackage[colorlinks,
linkcolor=red,
anchorcolor=blue,
citecolor=green
]{hyperref}

\ifCLASSINFOpdf
\else
\fi
\usepackage{amsmath}
\usepackage{diagbox}

\usepackage{ amssymb }
\usepackage{amsfonts}
\usepackage{caption}
\usepackage{graphicx}
\usepackage{lettrine}
\usepackage{epstopdf}
\usepackage{textcomp,booktabs}
\usepackage[usenames,dvipsnames]{color}
\usepackage{colortbl}
\definecolor{mygray}{gray}{.9}
\definecolor{mypink}{rgb}{.99,.91,.95}
\definecolor{mycyan}{cmyk}{.3,0,0,0}
\usepackage{pifont}
\usepackage{subfigure}
\usepackage{multirow}
\usepackage{url}
\usepackage{color}
\usepackage{threeparttable}
\usepackage{setspace}
\usepackage{lineno}
\usepackage{tabularx}
\usepackage[dvipsnames,usenames]{color}

\usepackage{bm}

\usepackage{algorithm} 
\usepackage{algorithmic} 

\usepackage{dblfloatfix}

\usepackage[dvipsnames,usenames]{color}
\usepackage[usenames,dvipsnames]{color}

\definecolor{light-gray}{gray}{0.90}
\begin{document}


	\title{ Semantic Communications  with World Models}

	\author{Peiwen Jiang,~\IEEEmembership{Member,~IEEE,} Jiajia Guo,~\IEEEmembership{Member,~IEEE,} Chao-Kai Wen,~\IEEEmembership{Fellow,~IEEE,}\\ Shi Jin,~\IEEEmembership{Fellow,~IEEE,} and Jun Zhang,~\IEEEmembership{Fellow,~IEEE}
			\thanks{P. Jiang, J. Guo and J. Zhang are with the Department of Electronic and Computer Engineering,
				Hong Kong University of Science and Technology, Hong Kong (e-mail: eepwjiang@ust.hk; eejiajiaguo@ust.hk; eejzhang@ust.hk). (Corresponding author: Jun Zhang.)}
			\thanks{C.-K. Wen is with the Institute of Communications Engineering, National
				Sun Yat-sen University, Kaohsiung 80424, Taiwan (e-mail: chaokai.wen@mail.nsysu.edu.tw).}
			\thanks{S. Jin is with the School
				of Information Science and Engineering, Southeast University, Nanjing
				210096, China (e-mail: jinshi@seu.edu.cn).}}
	
	\maketitle
	\pagestyle{empty}  
	\thispagestyle{empty} 

\begin{abstract}

Semantic communication is a promising technique for emerging wireless applications, which reduces transmission overhead by transmitting only task-relevant features instead of raw data. However, existing methods struggle under extremely low bandwidth and varying channel conditions, where corrupted or missing semantics lead to severe reconstruction errors. To resolve this difficulty, we propose a world foundation model (WFM)-aided semantic video transmission framework that leverages the predictive capability of WFMs to generate future frames based on the current frame and textual guidance. This design allows transmissions to be omitted when predictions remain reliable, thereby saving bandwidth.  Through WFM's prediction, the key semantics are preserved, yet minor prediction errors tend to amplify over time.  To mitigate issue, a lightweight depth-based feedback module is introduced to determine whether transmission of the current frame is needed. Apart from transmitting the entire frame, a segmentation-assisted partial transmission method is proposed to repair degraded frames, which can further balance performance and bandwidth cost. Furthermore, an active transmission strategy is developed for mobile scenarios by exploiting camera trajectory information and proactively scheduling transmissions before channel quality deteriorates.  Simulation results show that the proposed framework significantly reduces transmission overhead  while maintaining task performances across varying scenarios and channel conditions.

\end{abstract}


\section{Introduction}

\IEEEPARstart{S}{emantic}  communication is effective to reduce transmission overhead, which enables efficient information exchange by prioritizing meaning preservation over bit-level accuracy. This capability is enabled by several state-of-the-art techniques \cite{qin2021semantic, yang2022semantic, zhang2024intellicise}, including joint source-channel coding, shared knowledge base (KB), and task-aware semantic compression. However, these semantic techniques are only effective under specific scenarios. Thus, recent advancements \cite{ma2023task, zhang2024scan, wang2024adaptive, oh2025blind} have further refined these approaches by integrating context- and environment-aware adaptive coding and physical modules, where semantic relevance is dynamically assessed based on real-time task requirements and channel conditions. Among these novel semantic architectures, establishing KBs and utilizing generative models are becoming a research hotspot \cite{10628028, liang2024generative, 10554663,10318078}.  Generative models provide a flexible framework that can compensate for missing semantic features and reconstruct acceptable results.

Recently, the rise of generative foundation models (FMs), including large language models (LLMs), diffusion models (DMs), and vision-language models (VLMs), has further revolutionized this field by understanding context in a human-like manner. The large number of parameters and the generative architectures of FMs provide an unprecedentedly powerful KB and greatly enhance the ability to generate and complete content. Integrating FMs into semantic communication systems can drastically improve performance and adaptability \cite{jiang2023large}. FMs have been employed to coordinate multiple models \cite{shen2023large} and to control specific modules for different tasks \cite{wang2023seggpt, shao2024wirelessllm, xu2024semantic}. The general generative capability of FMs can be leveraged to restore transmitted semantic features even with lossy data \cite{grassucci2025lightweight}. Moreover, FMs are beneficial for exploiting potential KBs between the transmitter and receiver, which can further reduce transmission overhead. For example, the relationship between the image content and camera position  has been exploited to provide a strong pre-reconstruction basis at the receiver \cite{jiang2024position}. These FM-based approaches offer robust solutions for understanding complex transmission content and demonstrate remarkable adaptability in dynamic wireless communication environments.  
  
Among the proliferation of foundation models, world foundation models (WFMs) \cite{yang2024cogvideox, agarwal2025cosmos, bruce2024genie} are poised to revolutionize existing transmission paradigms by bridging the gap between digital representation and physical reality. Many downstream applications benefit from WFMs, including video generation, environment modeling, and decision making. Sora \cite{liu2024sora} can generate videos where objects maintain consistent appearances across different camera angles and lighting conditions, though it still struggles with complex physical interactions. Vista \cite{gao2024vista} provides high-fidelity and highly controllable video simulation for motion assessment. DreamerV3 \cite{hafner2023dreamerv3} utilizes WFMs to achieve efficient behavioral learning, yielding higher sample efficiency in robotics control tasks. Unlike conventional models constrained to pattern recognition within training datasets, WFMs exhibit an unprecedented capacity to comprehend fundamental physical laws. This intrinsic understanding enables them not only to perform reactive predictions but also to proactively simulate future states based on real-time environmental conditions.

In  video communications, frames demonstrate significant correlation in both temporal and spatial domains. Conventional methods attempt to leverage this correlation, but they still face numerous challenges. Traditional video coding primarily addresses pixel-level redundancy and is weak in predicting high-level semantic changes. Moreover, current semantic models have limited understanding and inadequate adaptability for capturing dynamics of physical objects. Inspired by the capability of WFMs, this study explores the temporal relationships among transmitted content to reduce transmission overhead from a new perspective. Unlike existing approaches that rely on pre-shared KBs, receiver caches, or periodic retransmissions, the proposed framework leverages the predictive capability of WFMs to omit certain frames from transmission. To ensure reliability, a flexible strategy is introduced that combines multiple transmission modes with a lightweight feedback mechanism, which adaptively triggers transmission only when prediction quality degrades. Furthermore, a semantic segmentation-assisted repair mechanism and a trajectory-aware active scheduling method are incorporated to mitigate the impact of poor channel conditions and avoid disconnections in advance. Ablation studies verify the effectiveness of introducing WFMs and highlight the potential of WFM-based semantic communication as these models continue to evolve.

The contributions of our study are as follows:
\begin{itemize}
\item \textbf{Novel transmission strategies using WFM prediction:} The proposed WFM-based strategy only requires occasional frame transmission because subsequent frames can be accurately predicted using the WFM based on the current frame and guidance from the motion state. Compared with conventional video coding and transmission methods, WFMs offer a better understanding of the physical world and provide more authentic results in bandwidth-limited scenarios. By omitting most transmissions, the proposed framework outperforms competing approaches in reducing overhead and improving performance.

\item \textbf{Flexible bandwidth cost with the feedback guidance:} The transmitted content and channel conditions affect prediction accuracy to some extent. For example, WFMs can predict more frames when the camera moves smoothly and the input frame is of high quality. To handle varying scenarios, a feedback mechanism is proposed using additional depth information, which requires little bandwidth to monitor prediction errors. When poor performance is detected, a new transmission is initiated, thereby enabling flexible bandwidth usage.

\item \textbf{Adaptability under varying channel conditions and disconnections:} When the camera trajectory is known in advance, transmission times can be proactively selected to avoid poor channel conditions. For instance, when the camera moves and the signal-to-noise ratio (SNR) decreases, frames should be transmitted earlier to prevent transmission under the worst conditions.
\end{itemize}

The remainder of this paper is organized as follows. Section \uppercase\expandafter{\romannumeral2} describes the system model, including the semantic transmission process and WFM prediction. The proposed framework is presented in Section \uppercase\expandafter{\romannumeral3}. Section \uppercase\expandafter{\romannumeral4} demonstrates the proposed networks can reduce the bandwidth cost and improve the transmission quality under various contents and channel conditions. Finally, Section \uppercase\expandafter{\romannumeral5} concludes the paper.

\section{System Model}
\label{SystemModel}
In this section, we introduce the classic semantic transmission framework and the conventional communication modules under a mobility scenario. Furthermore, the process of the WFM is described. Finally, the motivations and challenges of WFM-aided semantic communication are discussed.

\subsection{Semantic Transmission under A Wireless Communication System} \label{SemanticCommunication&OFDM}

We first present the basic semantic transmission process within a wireless communication system. At the transmitter, a semantic encoder $\Theta_{\rm en}$ compresses the input image $\mathbf{p}$ into a compact codeword by exploiting domain-specific KBs. The encoding process is expressed as
\begin{equation}
	\mathbf{b} = {\tt SC}_{\rm en}(\mathbf{p}; \Theta_{\rm en}),
\end{equation}
where ${\tt SC}_{\rm en}(\cdot)$ denotes the semantic compression function. The codeword $\mathbf{b}$ is subsequently modulated and mapped into an OFDM symbol $\mathbf{x}$ for transmission. 

During wireless transmission, the received frequency-domain signal is modeled as
\begin{equation}
	\mathbf{y} = \mathbf{h} \odot \mathbf{x} + \mathbf{z},
\end{equation}
where $\mathbf{h} \in \mathbb{C}^{K \times 1}$ denotes the channel frequency response across $K$ subcarriers, and $\mathbf{z}$ represents additive white Gaussian noise (AWGN). At the receiver, channel estimation provides $\widehat{\mathbf{h}}$, and equalization yields
\begin{equation}
	\widehat{\mathbf{x}} = \mathbf{y} \oslash \widehat{\mathbf{h}},
\end{equation}
where $\oslash$ denotes element-wise division. After demodulation, the estimated codeword $\widehat{\mathbf{b}}$ is obtained and passed to the semantic decoder:
\begin{equation}
	\widehat{\mathbf{p}} = {\tt SC}_{\rm de}(\widehat{\mathbf{b}}; \Theta_{\rm de}),
\end{equation}
where ${\tt SC}_{\rm de}(\cdot)$ uses receiver-side KBs $\Theta_{\rm de}$ to reconstruct the image $\widehat{\mathbf{p}}$.  
This end-to-end design highlights how semantic-aware processing improves communication efficiency by transmitting feature-level information rather than raw pixels.

In addition to fixed-SNR evaluations commonly adopted in existing semantic communication studies, our experiments also consider a moving scenario. An uplink cellular network is modeled with two base stations (BSs) deployed 500 m apart along the x-axis. The BSs are located at $\text{BS}_1 = (-250, 0)$ m and $\text{BS}_2 = (250, 0)$ m within a 1,000 $\times$ 1,000 m area. Mobile users are uniformly distributed at a height of $h_r = 1.5$ m, while BS antennas are mounted at $h_t = 10$ m.

The large-scale path loss is modeled using the modified COST 231 Hata model for urban environments  
\cite{hata2013empirical}:
\begin{equation}
	\begin{aligned}
		L(d) = &46.3 + 33.9\log_{10}{\left(\frac{f}{1\text{ MHz}}\right)} - 13.82\log_{10}(h_t) \\
		&- a(h_r) + (44.9 - 6.55\log_{10}(h_t))\log_{10}{\left(\frac{d}{1\text{ km}}\right)},
	\end{aligned}
\end{equation}
where $f = 2$ GHz is the carrier frequency, $d$ is the transmitter-receiver distance in meters, and $a(h_r)$ is the mobile station correction factor:
\begin{equation}
	a(h_r) = 3.2\left[\log_{10}(11.75h_r)\right]^2 - 4.97.
\end{equation}
The received power at position $(x,y)$ from BS$_i$ is given by
\begin{equation}
	P_{r}(x,y) = P_t + G - L(d(x,y)), \label{PG}
\end{equation}
where $P_t$ dBm is the transmit power, $G$ dBi is the antenna gain, and $d$ is the distance to the BS.  
The corresponding  SNR is expressed as
\begin{equation}
	\text{SNR}(x,y) = 10\log_{10}\left(\frac{P_{r}(x,y)}{ N_0}\right),
\end{equation}
where $N_0 = -174 + 10\log_{10}(B)$ dBm denotes the thermal noise power with system bandwidth $B = 20$ MHz.

\begin{figure*}
	\centering
	{\includegraphics[width=0.9\linewidth]{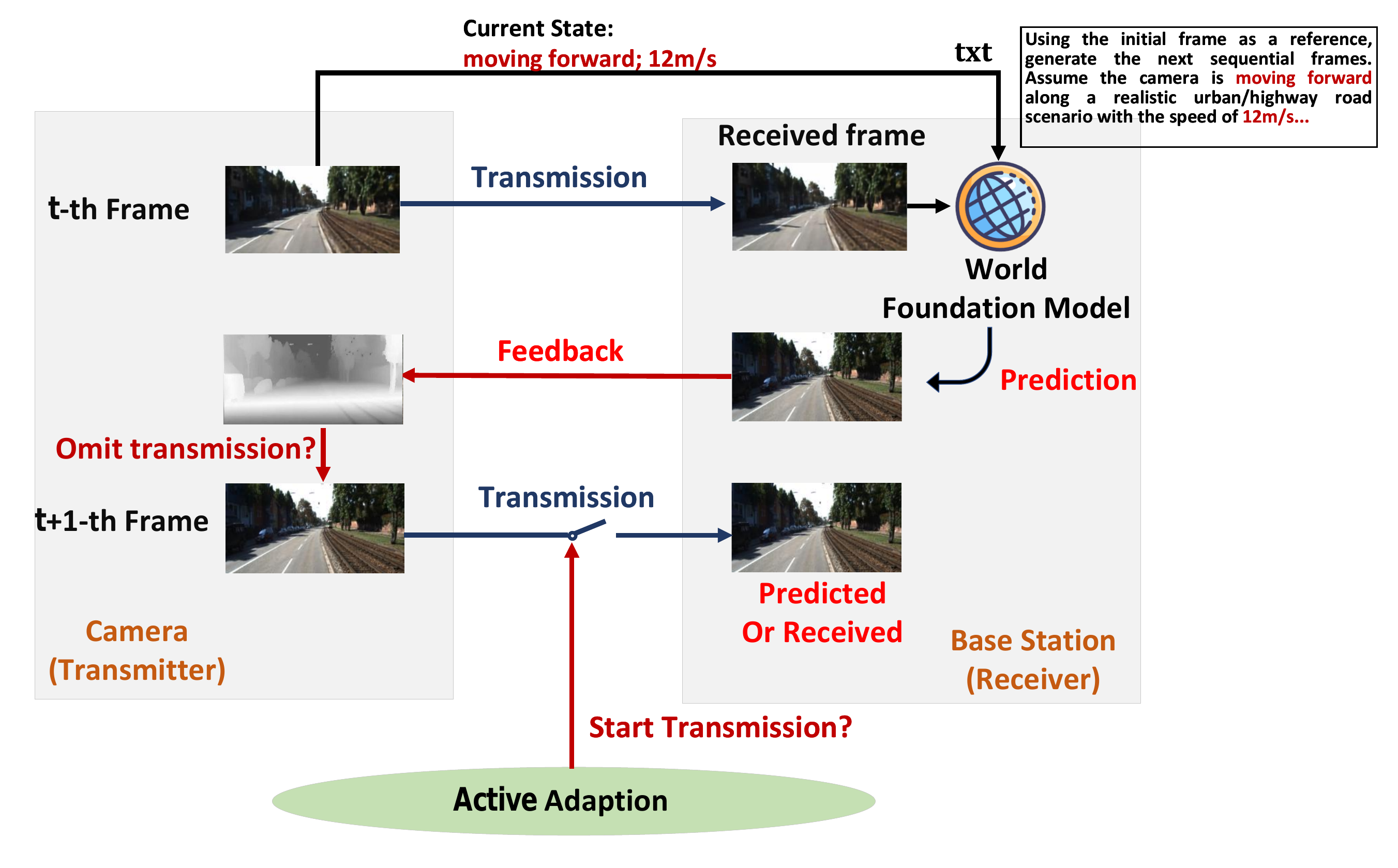}}
	\caption{Overview of the proposed framework.}
	\label{Overall}
\end{figure*}

\subsection{World Foundation Models}
WFMs are advanced AI systems designed to simulate and understand the dynamics of the physical world. Their rapid evolution in recent years has been driven by breakthroughs in video generation, synthetic data creation, and multimodal reasoning, as exemplified by OpenAI's Sora \cite{liu2024sora}, Zhipu AI's CogVideoX \cite{yang2024cogvideox}, and NVIDIA's Cosmos \cite{agarwal2025cosmos}.

A typical WFM, such as Sora, consists of several key modules, including a visual encoder-decoder, a text encoder, and a Transformer-based diffusion model (DiT). The visual encoder-decoder, such as ViViT \cite{arnab2021vivit} or 3D VAE \cite{yu2023language}, partitions the input frames into 3D patches that capture both spatial and temporal dimensions. This process is expressed as
\begin{equation}
\mathbf{Z}={\tt V}_{\rm en}([\mathbf{p}_{0},\ldots,\mathbf{p}_{T}]),
\end{equation}
where $\mathbf{Z}$ is the patch matrix with height, width, spatial, and temporal dimensions, and $[\mathbf{p}_{0}, \ldots, \mathbf{p}_{T}]$ denotes the set of input frames. By modeling these patches, the WFM extracts spatiotemporal features to ensure coherent action representation.
The text encoder, such as CLIP \cite{radford2021learning}, converts the input text $\mathbf{txt}$ into an embedding to guide the generative process:
\begin{equation}
\mathbf{c} = {\tt T}_{\rm en}(\mathbf{txt}).
\end{equation}
The DiT receives both visual and text embeddings to generate continuous frame sequences. Similar to conventional diffusion models, $\mathbf{Z}$ is perturbed with noise and then progressively denoised to obtain $\tilde{\mathbf{Z}}$. Frame length and resolution can also be controlled by adjusting the patch representations. For convenience, the process can be summarized as
\begin{equation}
	\tilde{\mathbf{Z}}={\tt DiT}(\mathbf{Z}, \mathbf{c}).
\end{equation}
Finally, the generated frames are reconstructed through the visual decoder:
\begin{equation} 
[\tilde{\mathbf{p}}_{0}, \ldots, \tilde{\mathbf{p}}_{T}] = {\tt V}_{\rm de}(\tilde{\mathbf{Z}}).
\end{equation}

With this architecture, WFMs support diverse applications such as text-to-video, image-to-video, and video-to-video generation. Compared with conventional generative foundation models, WFMs provide a stronger understanding of physical dynamics in both spatial and temporal domains, making them valuable as data generators for training other models, such as autonomous driving systems. In this study, WFMs are employed to predict future transmission content by leveraging the current image and textual guidance:
\begin{equation}
[\mathbf{p}_{0}, \ldots, \tilde{\mathbf{p}}_{T}] = {\tt WFM}(\mathbf{p}_{0}, \mathbf{txt}), \label{WFMeq}
\end{equation}
where ${\tt WFM}(\cdot)$ represents the entire process described above.

\subsection{Benefits and Challenges of Integrating WFMs}
WFMs capture causal, temporal, and physical regularities that allow semantic communication to compress information toward task-relevant world states rather than surface-level representations. This capability enables goal-oriented and bandwidth-efficient communication through shared multimodal priors and the prediction of downstream consequences. Compared with existing FM-based semantic methods, WFMs further support cross-situational alignment, continual adaptation to dynamic environments, and policy-conditioned semantics.

Despite these advantages, integrating WFMs also introduces several challenges. First, the generative dynamics of WFMs may propagate bias from synthetic results, potentially degrading communication accuracy. Second, the substantial computational requirements impose stress on edge devices and real-time communication links. Finally, evaluation and control strategies are critical, as the focus shifts from transmission accuracy to task success under uncertainty. This shift demands the development of new datasets, simulation platforms, and verification tools to ensure robust and reliable deployment of WFM-aided semantic communication.

\section{Semantic Communications with World Foundation Models}
\label{s3}

This section presents the proposed framework, which fully exploits the capability of WFMs at the BS. To improve reconstruction performance and cope with varying channel conditions, several adaptive methods are introduced in cooperation with different foundation models. Their network architectures and workflows are described in detail. Finally, the camera trajectory is incorporated into the framework design, where active transmission strategies are recommended.

\subsection{General Workflow}

The general framework is illustrated in Fig.~\ref{Overall}, where the WFM operates at the receiver (BS) and predicts subsequent frames based on the transmitted image and the prompt text. A monitoring mechanism determines whether the next frame should be transmitted or omitted, requiring an appropriately designed encoder-decoder. Unlike conventional semantic communication methods, transmission is not required at every timeslot, which significantly reduces bandwidth overhead and maintains operation even under temporary communication loss. 
The workflow of the proposed framework consists of four main components:
\begin{itemize}  
	
        \item \textbf{WFM's Prediction:} The WFM uses the current image as a beginning, capturing details like object positions, shapes, and scene context, while the prompt text offers guiding instructions, be it specifying object actions or introducing new elements.  Then, it generates subsequent frames that are visually consistent with the current image and adhere to the prompt, and may even refine the output to ensure high quality and coherence.
        
    \item \textbf{Optional Transmission:} The first frame must be fully transmitted because an accurate image is essential for reliable WFM prediction. Accordingly, an encoder-decoder is designed for this transmission, which requires sufficient bandwidth and favorable channel conditions. As the channel environment varies with camera movement, full transmission may not always be feasible when predicted frames degrade. In such cases, key semantic features can be transmitted instead, allowing the receiver to repair degraded frames.

	\item \textbf{Depth Feedback:} When transmission is omitted, the receiver cannot access the true frame and prediction quality degrades over time. To address this, a low-cost monitor is deployed at the receiver, and depth maps are fed back to the transmitter. Based on this feedback and current channel conditions, the transmitter determines whether to initiate a new transmission and selects the appropriate encoder-decoder.
	
	\item \textbf{Adaptation:}  Some foundation models, such as the Segment Anything Model (SAM) \cite{wang2023seggpt} and diffusion models, are employed to reconstruct degraded frames. Besides, the transmission time also affects  the following frames. Since performance is influenced by both transmitted content and channel conditions, an adaptive strategy is required to coordinate these architectures effectively.
\end{itemize}

\begin{figure*}
	\centering
\subfigure[]{\includegraphics[width=0.9\linewidth]{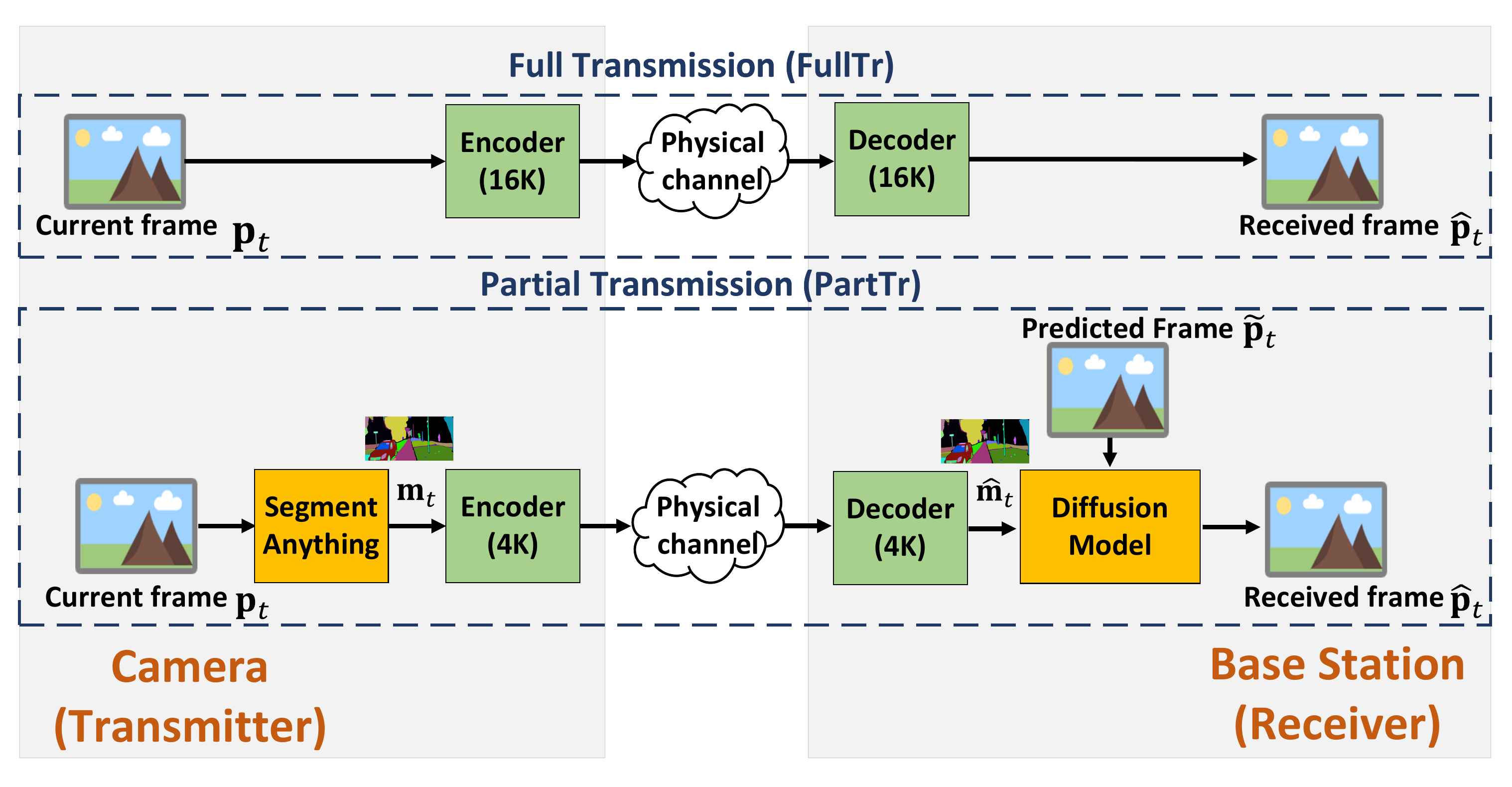}}
\subfigure[]{\includegraphics[width=0.8\linewidth]{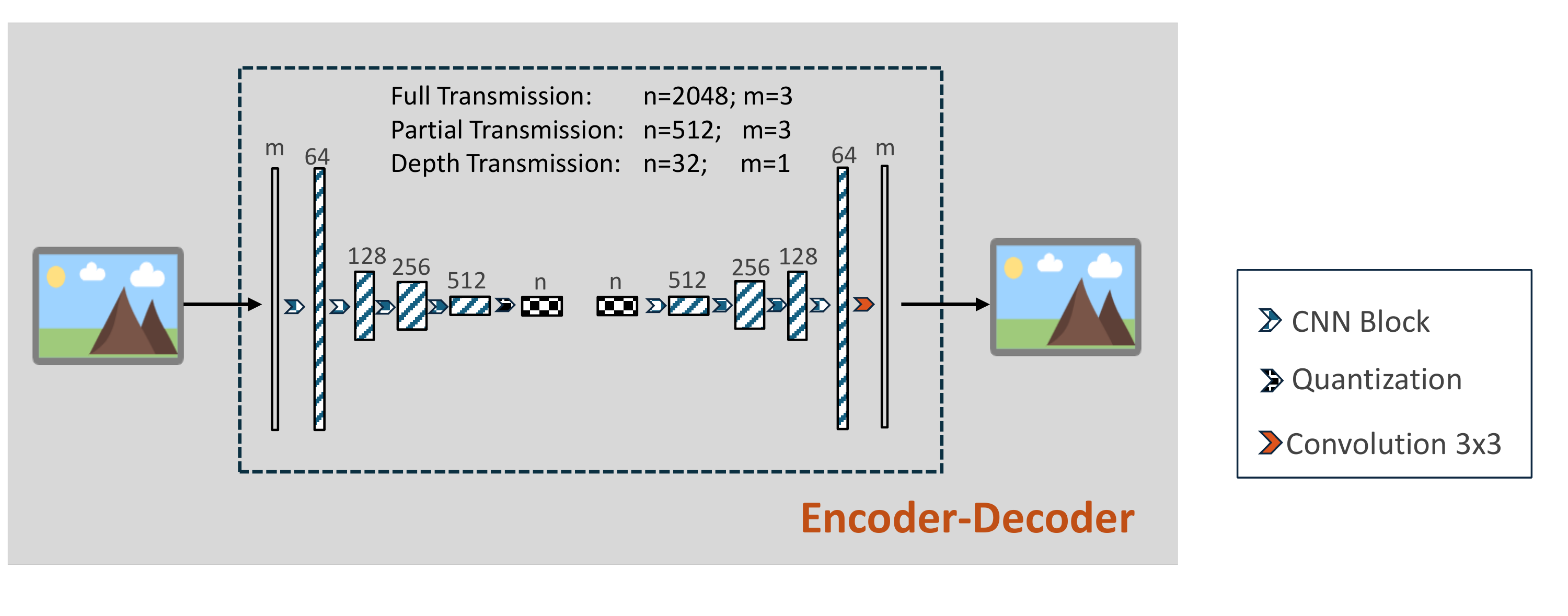}}

	\caption{(a) Architectures of two different transmission methods. (b) Detailed network of the related encoder and decoder.} 
	\label{Detailed network}
\end{figure*}

In this framework, the received frame at the $t$-th timeslot can be obtained through one of three approaches: full transmission (FullTr), WFM's prediction, or partial transmission (PartTr).  Conventional  Hybrid Automatic Repeat reQuest (HARQ) improves reliability by requesting additional parity when a decoded packet fails, which remedially protects past transmissions at the codeword level. In contrast, our WFM-aided framework issues predictive semantic requests. The receiver first synthesizes future frames with the WFM and continues without transmission while a lightweight depth-based monitor indicates sufficient quality. When the monitor predicts quality degradation, it requests minimal semantic redundancy (PartTr) or, if necessary, a FullTr.
The detailed architectures implementing these methods are presented in the following subsections. 

\subsection{WFM's Prediction}
Leveraging WFMs to predict future frames is a core component of the proposed framework. Here, the pre-trained CogVideoX \cite{yang2024cogvideox} is employed, and the process follows (\ref{WFMeq}). As shown in Fig.~\ref{Overall}, the input text of the WFM is given as 
\textit{``Using the initial frame as a reference, generate the next sequential frames. Assume the camera is moving forward along a realistic urban or highway scenario with a speed of 12 m/s...''}, 
where the camera state is updated continuously. Since updating text prompts consumes far fewer resources than image transmission, text transmission is omitted from this study. The input image, however, is subject to channel noise, and therefore reliable image transmission provides an essential initialization. Suppose the image at the $t$-th timeslot $\hat{\mathbf{p}}_{t}$ is received, the following frames can be predicted as 
\begin{equation}
[\hat{\mathbf{p}}_{t}, \tilde{\mathbf{p}}_{t+1}, \ldots, \tilde{\mathbf{p}}_{t+T}] = {\tt WFM}(\hat{\mathbf{p}}_{t}, \mathbf{txt}),
\end{equation}
where $\mathbf{txt}$ describes the camera state and guides the generation of the next $T$ frames.

\subsection{Optional Transmission }

\subsubsection{Full Transmission (FullTr)}
As shown at the top of Fig.~\ref{Detailed network}(a), the FullTr method transmits the entire image directly. This approach requires the largest bandwidth among all strategies and is highly sensitive to channel variations. The basic encoder-decoder structure, illustrated in Fig.~\ref{Detailed network}(b), is composed of several convolutional neural network (CNN) blocks. The overall process can be expressed as
\begin{equation}
	\hat{\mathbf{p}}_{t}={\tt SC}_{\rm de, full}(h({\tt SC}_{\rm en, full}(\mathbf{p}_{t}))), \label{basic}
\end{equation}
where ${\tt SC}_{\rm en, full}(\cdot)$ and ${\tt SC}_{\rm de, full}(\cdot)$ denote the encoder and decoder for FullTr, respectively, and $h(\cdot)$ represents the effect of the physical channel.

\begin{itemize}
    \item \textbf{${\tt SC}_{\rm en, full}$:} The encoder begins with a $13 \times 13$ convolution layer with 64 channels, followed by a $4 \times$ downsampling layer. Subsequent blocks use $7 \times 7$, $5 \times 5$, and $3 \times 3$ convolution layers with 128, 256, and 512 channels, each with $2 \times$ downsampling. ReLU activations are applied in all blocks. A quantization layer flattens the outputs, followed by a dense layer with Tanh activation, which produces a number of values equal to the transmitted bits. A hard-decision criterion converts these values into the binary sequence $\mathbf{b}$, with gradients rewritten for end-to-end training.  

\item \textbf{${\tt SC}_{\rm de, full}$:} The decoder mirrors the encoder, using four convolutional blocks with upsampling layers. A final $3 \times 3$ convolutional layer with Tanh activation reconstructs the frame $\widehat{\mathbf{p}}_{t}$.  

\item \textbf{$h(\cdot)$:} The binary sequence is mapped to 16-QAM symbols and transmitted through a standard OFDM transceiver. The SNR of the physical channel affects the received sequence and thus the final reconstruction. 

\end{itemize}

\begin{figure*}[!t]
	\centering
		{\includegraphics[width=1\linewidth]{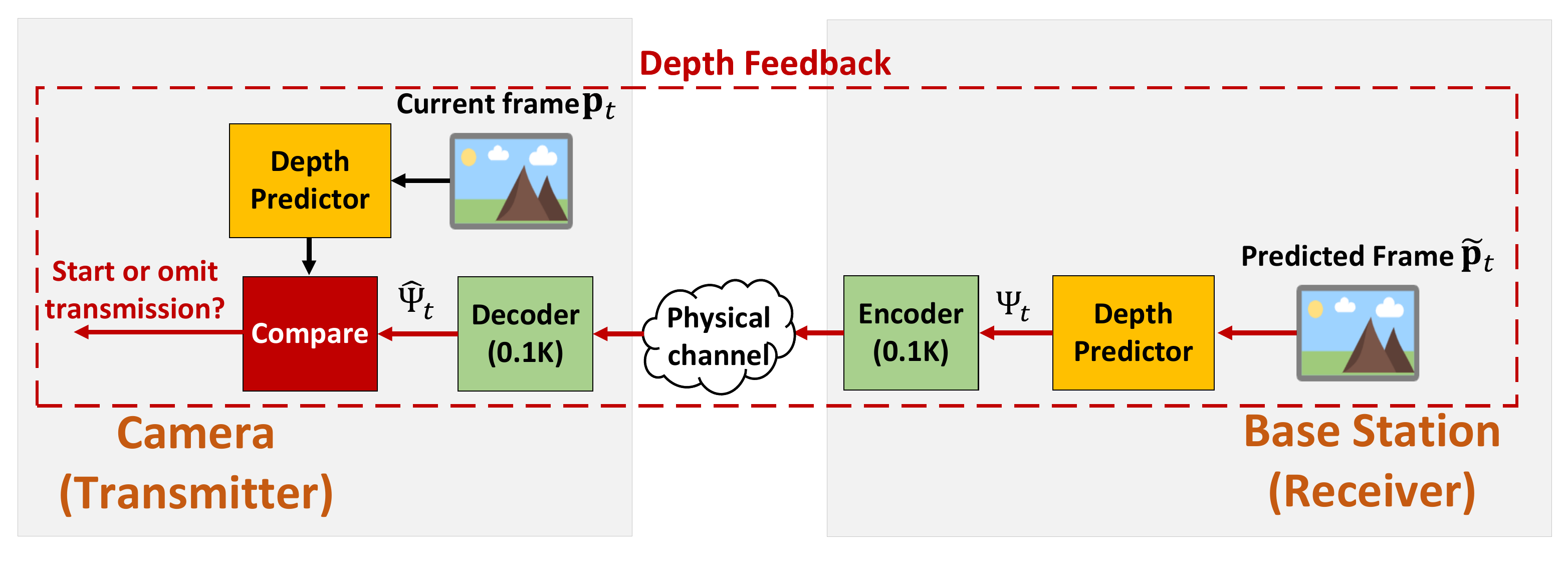}}
	\caption{Architecture of the proposed depth feedback, where the current depth map is fed back to decide whether transmission is required.} 
	\label{Monitorarchi}
\end{figure*}
\subsubsection{Partial Transmission (PartTr)}
Due to camera movement and varying channel conditions, full transmission may not always be feasible. When the WFM’s prediction quality degrades, partial transmission becomes necessary. As shown in Fig.~\ref{Detailed network}(a), PartTr integrates two FMs to enable efficient transmission and robust reconstruction.

At the transmitter, the pre-trained SAM \cite{wang2023seggpt} is applied to convert the original image into a semantic mask:
\begin{equation}
	\mathbf{m}_{t} = {\tt SAM}(\mathbf{p}_{t}),
\end{equation}
where ${\tt SAM}(\cdot)$ denotes the segmentation process and $\mathbf{m}_{t}$ is the mask, with different colors indicating distinct objects. This mask is transmitted using a similar encoder-decoder structure to (\ref{basic}), but with reduced complexity:
\begin{equation}
	\hat{\mathbf{m}}_{t} = {\tt SC}_{\rm de, part}\big(h({\tt SC}_{\rm en, part}(\mathbf{m}_{t}))\big),
\end{equation}
where ${\tt SC}_{\rm en, part}(\cdot)$ and ${\tt SC}_{\rm de, part}(\cdot)$ denote the encoder and decoder for PartTr, respectively. Since the mask omits fine image details, the encoder output dimension is reduced to 512, requiring only one quarter of the bandwidth compared with FullTr. The decoder is modified accordingly.

At the receiver, the reconstructed mask $\hat{\mathbf{m}}_{t}$ is combined with the WFM-predicted frame $\tilde{\mathbf{p}}_{t}$. A conditional DM is retrained to generate the repaired frame:
\begin{equation} 
	\hat{\mathbf{p}}_{t} = {\tt DM}(\hat{\mathbf{m}}_{t}, \tilde{\mathbf{p}}_{t}),
\end{equation}
where ${\tt DM}(\cdot)$ is trained on triplets $(\tilde{\mathbf{p}}_{t}, \hat{\mathbf{m}}_{t}, \mathbf{p}_{t})$ under varying SNRs and timeslots. This design enables robust recovery of degraded frames while significantly reducing bandwidth consumption.

\subsection{Depth Feedback}
\label{s3b}

Prediction quality inevitably degrades over time, and transmission becomes necessary. To decide when to transmit, the predicted frame must be monitored at the receiver, which then informs the transmitter. The proposed feedback mechanism sends the depth map of the next predicted frame to the transmitter as shown in Fig. \ref{Monitorarchi}. Specifically, the receiver extracts the depth map of $\tilde{\mathbf{p}}_{t+1}$:
\begin{equation} 
\mathbf{\Psi}_{t} = {\tt Dep}(\tilde{\mathbf{p}}_{t}),
\end{equation}
where $\mathbf{\Psi}_{t+1}$ denotes the depth map obtained from a pre-trained depth estimator \cite{yang2024depth}. This map is transmitted to the transmitter through a lightweight encoder-decoder similar to Fig.~\ref{Detailed network}(b), but requiring only $1/64$ of the bandwidth. When the true frame $\mathbf{p}_{t+1}$ is captured, the transmitter compares the estimated depth map with the feedback map:
\begin{equation}
\text{Transmission} = \left\{
\begin{aligned}
	&\text{yes}, \quad \delta_{1.25}({\tt Dep}(\mathbf{p}_{t}), \hat{\mathbf{\Psi}}_{t}) > \sigma, \\
	&\text{no}, \quad \text{otherwise},
\end{aligned}\right.
\end{equation}
where $\delta_{1.25}(\cdot)$ is the comparison criterion, $\hat{\mathbf{\Psi}}_{t}$ is the feedback depth map at the transmitter, and $\sigma = 0.3$ is the threshold.

In summary, the proposed predictor and monitor allow future frames to be directly predicted until transmission is reactivated. Once a frame is transmitted, it replaces the degraded predicted frame and serves as the new initialization for WFM prediction.

\begin{figure*}[!t]
	\centering

		{\includegraphics[width=1\linewidth]{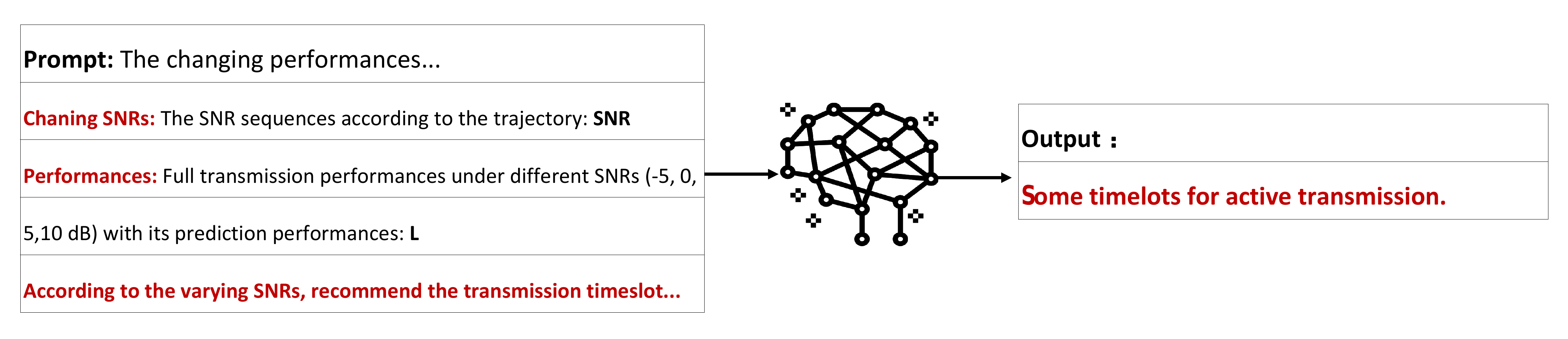}}
	\caption{ LLM-based active strategy generation.} 
	\label{LLM}
\end{figure*}

\subsection{Active Adaption}

Although the proposed framework can reduce transmission overhead through WFM prediction, it still relies on passive adaptation, where the transmission method is selected according to channel conditions and current performance. In practical vehicular scenarios, however, the trajectory of the camera is often known. This knowledge can be exploited to proactively determine transmission timing and strategy.


Unlike the depth feedback, which reacts to degraded prediction quality, the active strategy anticipates future channel conditions and triggers transmission before the SNR deteriorates. Otherwise, if triggered solely by the feedback, the optimal transmission time may be missed. In traditional form, this problem is usually solved by proposing a good and explicit rule to take the current situation into account and output an optimized solution. However, different rules should be considered under different scenarios and transmitted contents. Besides, when the adaptive modules, such as optional transmission methods, are increased, the strategy should be changed.

To simplify this problem, an LLM is used to generate an active strategy.  That means the LLM will combine the requirements and the current conditions, and then recommend some proper timeslots for active transmissions. For different tasks and scenarios, a unified input and output guidelines is adopted. An example is described as follow:

1) The SNR sequence is predicted from the vehicle’s distance to BSs:
\begin{equation}
	\mathbf{SNR}=[SNR_0, SNR_1, \ldots, SNR_T]
\end{equation}
where $\mathbf{SNR}$ denotes the estimated values for the following $T$ timeslots. 

2) The degradation of WFM prediction is modeled as
\begin{equation}
	\mathbf{L}=[l_0, \ldots, l_T],
\end{equation} 
where $\mathbf{L}$ is estimated from the transmission content.  In this study, three scenarios are evaluated: a basic street, a busy street, and a crossroad. The average performance degradation curves obtained from these scenarios are used as the corresponding $\mathbf{L}$ references. 

To generate an active strategy, the SNR and degradation sequences are provided to an LLM, which outputs decisions for selecting appropriate timeslots. These decisions are executed in collaboration with the depth-based feedback mechanism. The principle is straightforward: while the depth feedback only concerns the current timeslot, the LLM proactively initiates transmission even when the feedback does not request it. The transmission method (FullT or PartT) is then selected according to the current SNR and available bandwidth. 

Overall, the proposed WFM-based transmission framework significantly reduces transmission overhead. To ensure reliable performance in dynamic environments, a depth-based monitor and an LLM-driven active strategy are integrated, providing practical and efficient solutions. 


\section{Simulation Results}
\label{s5}
This section presents the simulation results of the proposed framework. After introducing the simulation parameters, we first evaluate the prediction performance of WFM with FullTr and PartTr under different scenarios. Next, we demonstrate the effectiveness of the proposed monitor. The active strategy is then tested as a superior choice when the camera trajectory is known. Finally, an ablation study is conducted to further assess the potential of the WFM-based transmission framework.

\subsection{Simulation Settings}

The KITTI dataset \cite{geiger2012we} is adopted, which contains more than ten thousand images captured from diverse areas, along with corresponding metadata such as timestamps and velocities. The dataset is divided into training and testing sets with a ratio of 10:1. All camera-view frames are cropped to $256 \times 128$, and every 20 consecutive frames are grouped into one set. The SAM is employed to generate segmentation masks for all images, with different objects labeled using different colors.

For comparison, the conventional baseline consists of AV1 video coding combined with LDPC channel coding. The baseline bandwidth consumption for 20 frames is approximately 80 KBytes with an LDPC code rate of 1/2. In contrast, FullTr requires 2 KBytes per frame, PartTr requires 0.5 KBytes per frame, and depth transmission requires 0.1 KBytes per frame. Consequently, transmitting 20 frames entirely with FullTr consumes 40 KBytes.  

In addition to the conventional PSNR metric, four modern metrics are introduced to evaluate performance across different tasks:  
\begin{itemize}
	\item \textbf{LPIPS:} Learned Perceptual Image Patch Similarity (LPIPS) is widely used as a perceptual similarity metric in computer vision. It computes the sum of mean squared errors (MSEs) between estimated and ground-truth images across different layers of a pretrained network, such as VGG, thereby reflecting feature similarity \cite{zhang2018unreasonable}.  

	\item \textbf{mIoU:} Mean Intersection over Union (mIoU) is a standard metric for semantic segmentation, evaluating the overlap between predicted segmentation maps and ground truth. It is computed as the average IoU across all classes.  

	\item \boldmath{$\delta > 1.25$:} This metric measures the percentage of pixels for which the relative error between predicted and ground-truth depth is within a threshold, commonly set to 1.25 \cite{ranftl2021vision}.  

	\item \textbf{DreamSim:} DreamSim \cite{fu2023dreamsim} is a perceptual similarity measure that integrates both low-level visual features and high-level semantics. It leverages CLIP, OpenCLIP, and DINO embeddings via contrastive learning, outperforming traditional metrics such as LPIPS in capturing fine perceptual and semantic differences.  
\end{itemize}

\subsection{Performance of Transmission and Prediction}
To evaluate the transmission and prediction performance of the proposed framework, different scenarios and channel conditions are compared. Each timeslot in the experiment corresponds to 0.5 s, and the fixed transmission interval is set to six timeslots. This means that the WFM is required to predict subsequent frames for approximately three seconds.

\begin{figure}[h]
	\centering
	\subfigure[]{		
		{\includegraphics[width=0.9\linewidth]{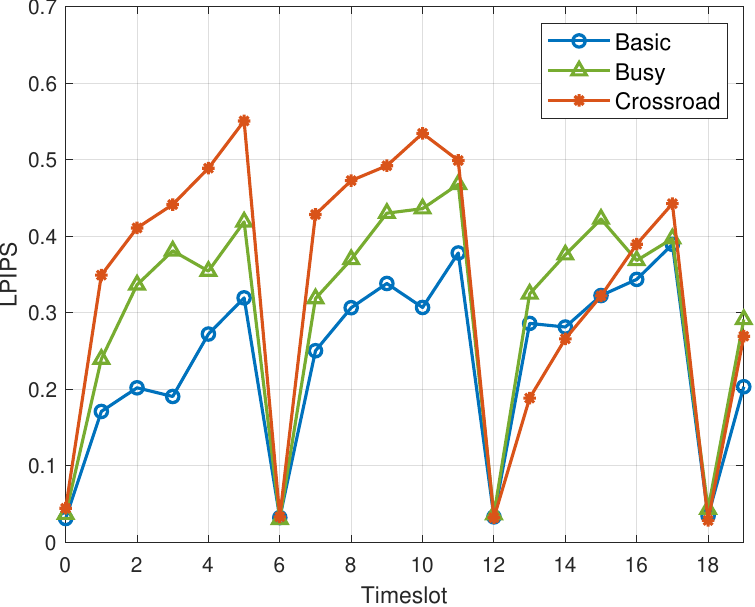}}}\\
	\subfigure[]{
		{\includegraphics[width=0.9\linewidth]{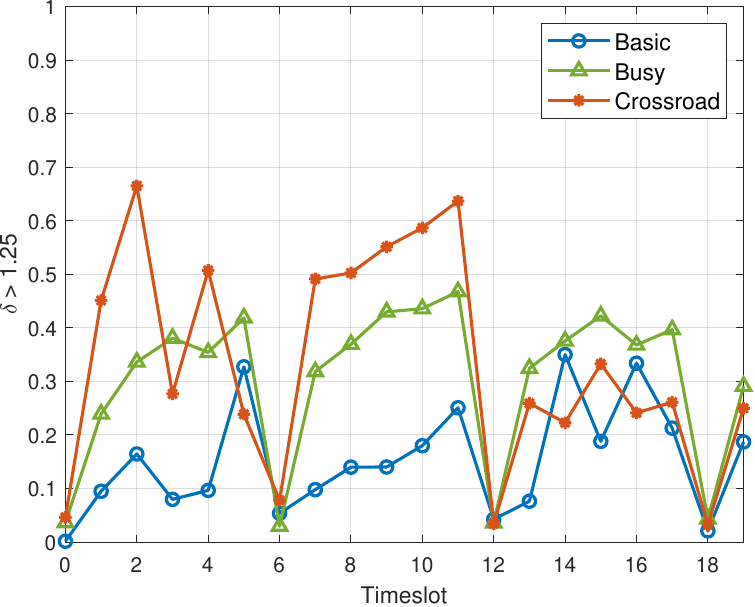}}}
	\caption{Performances of the WFM-based prediction under different transmission content with a fixed interval. (a) LPIPS metric. (b) Depth metric.}
	\label{SecA_sce}
\end{figure}
\begin{figure}[h]
	\centering
	\subfigure[]{
		
		{\includegraphics[width=0.9\linewidth]{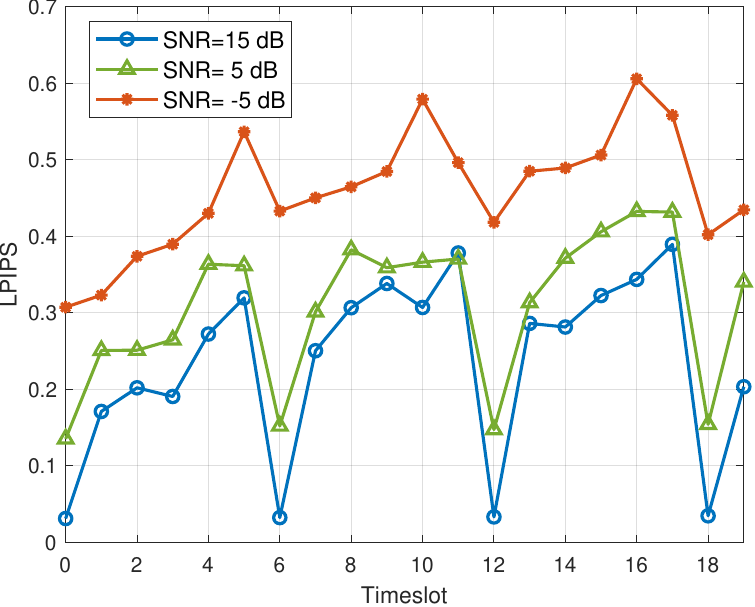}}}\\
	\subfigure[]{
		{\includegraphics[width=0.9\linewidth]{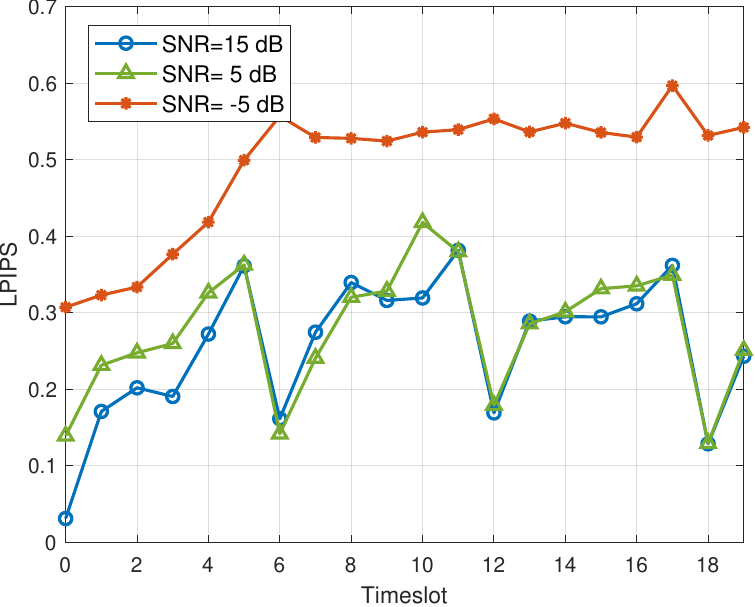}}}
	\caption{Performances of the WFM's prediction   under  varying transmission SNRs. (a) LPIPS metric. (b) Depth metric.}
	\label{SecA_SNR}
\end{figure}

\begin{figure*} 
	\centering
	
	\subfigure[Target]{
		\includegraphics[width=0.3\linewidth]{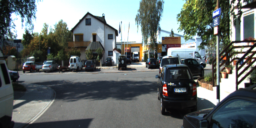}}
	\subfigure[PartTr]{
		\includegraphics[width=0.3\linewidth]{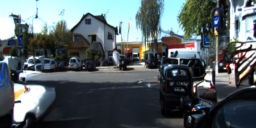}}
	\subfigure[Predicted]{
		\includegraphics[width=0.3\linewidth]{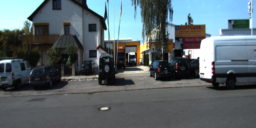}}
	\\
	
	\subfigure[Target]{
		\includegraphics[width=0.3\linewidth]{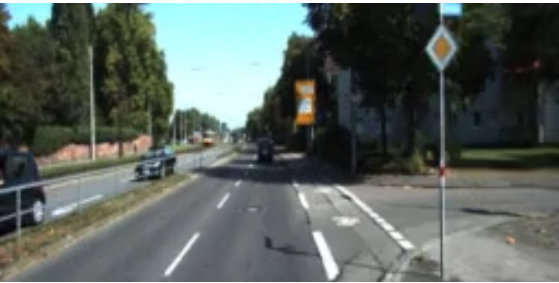}}
	\subfigure[FullTr]{
		\includegraphics[width=0.3\linewidth]{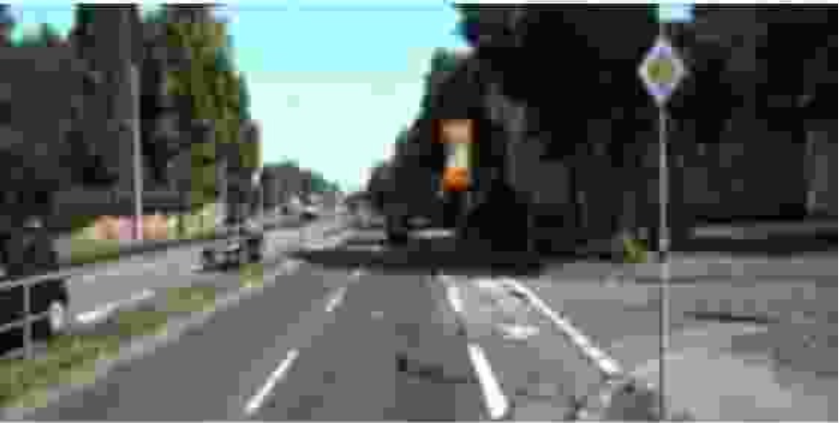}}
	\subfigure[Predicted]{
		\includegraphics[width=0.3\linewidth]{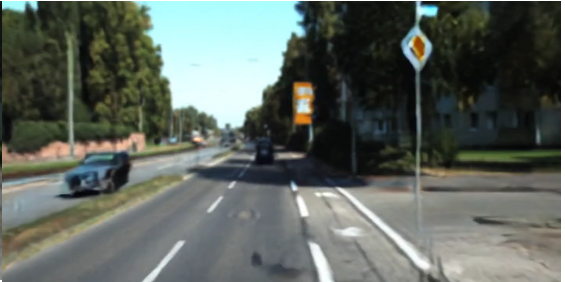}}
	\caption{Examples of the different transmission methods under SNR= 5 dB. The predicted frames are different from the target ones with the time going. Thus, the ParT method repairs the errors in the predicted frame while the FullTr method directly transmit the full target image.}
	\label{EX_A}
\end{figure*}

As shown in Fig.~\ref{SecA_sce}(a), the LPIPS performance across three typical scenarios is compared. In the \textit{basic} scenario, only a few vehicles are moving, making it relatively easy for the WFM to predict, and the performance degradation is the slowest. The \textit{busy street} scenario involves more moving vehicles, which are harder to predict, resulting in faster degradation. The \textit{crossroad} scenario is the most challenging due to camera turning, which causes rapid background changes. In this case, the prediction quality deteriorates quickly before the 12th timeslot, after which the camera completes the turn and performance stabilizes at a level comparable to the other scenarios. Fig.~\ref{SecA_sce}(b) presents the depth-based metric, which shows a similar trend but exhibits more instability when prediction quality is poor, since errors in transmission and prediction also affect the depth estimation. The basic scenario consistently achieves the best depth performance.

Fig.~\ref{SecA_SNR}(a) illustrates the LPIPS performance of the FullTr method under a fixed interval while the SNR varies. With a transmission interval of six, performance peaks at timeslots 0, 6, 12, and 18, then decreases in between. Interestingly, higher SNR values lead to lower LPIPS scores, indicating that prediction quality is also influenced by channel conditions, as transmission errors may mislead the WFM.

The performance of the PartTr method is shown in Fig.~\ref{SecA_SNR}(b). Here, FullTr is applied at the beginning, followed by PartTr for subsequent transmissions. Similar to FullTr, the performance at timeslot 0 is affected by SNR.  At timeslots 6, 12, and 18, the transmitted masks enable the PartTr method to repair the predicted frames. At high SNR (15 dB), the repaired frames are slightly inferior to FullTr because partial information cannot restore all details. At moderate SNR (5 dB), PartTr achieves performance comparable to FullTr. At very low SNR ($-5$ dB), PartTr fails to repair the degraded frames. Overall, PartTr effectively repairs WFM prediction errors and saves bandwidth, especially under moderate SNR conditions. However, due to its limited correction capability, the choice of transmission method and interval must remain flexible.

  \begin{figure*}[!t]
	\centering
	\subfigure[]{
		
		{\includegraphics[width=0.47\linewidth]{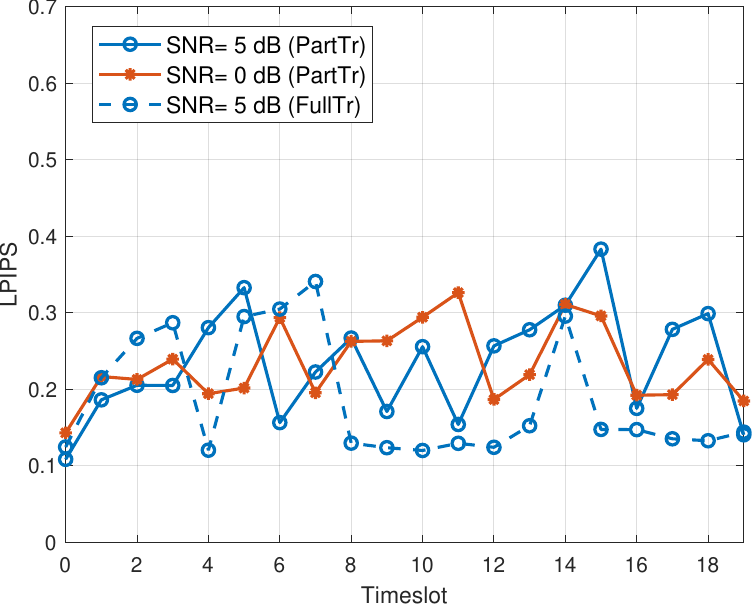}}}
	\subfigure[]{
		{\includegraphics[width=0.47\linewidth]{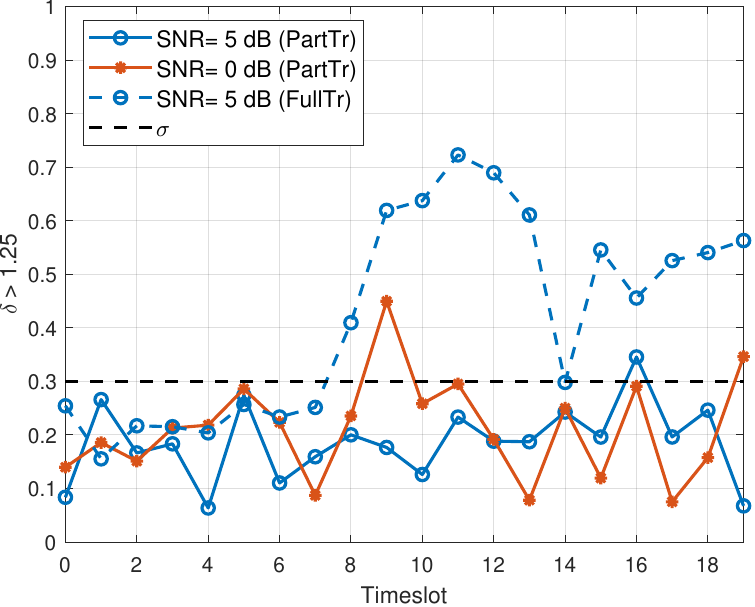}}}
	\caption{Performance of the competing methods using the proposed depth feedback. (a) LPIPS performance. (b) Threshold and corresponding depth performance.}
	\label{SecB_per}
\end{figure*}

Fig.~\ref{EX_A} provides visual examples of different transmission methods. In Fig.~\ref{EX_A}(a--c), predicted frames gradually deviate from the target, with objects such as vehicles and buildings moving closer to the camera. By applying PartTr, the predicted frame is repaired by the diffusion model, showing the effectiveness of the proposed design. Although some minor details are lost, the overall quality is significantly improved. In Fig.~\ref{EX_A}(d--f), when the predicted frame remains accurate and the SNR is 0 dB, the superiority of the WFM-based prediction is clear. While fully transmitted frames are degraded by noise, predicted frames maintain good perceptual quality without consuming transmission resources. In general, generation-based methods provide strong visual quality, though fine details may be lost, which highlights the importance of controlling accumulated errors.  

In summary, the combination of WFM-based prediction with FullTr and PartTr provides an effective means to reduce bandwidth consumption while adapting to varying channel conditions. Nevertheless, these gains depend on appropriate strategies that account for both the communication scenario and the prevailing channel conditions. To further control prediction errors in dynamic environments, the feedback mechanism is investigated in the following subsection.

\subsection{Performance With Depth Feedback}

In this subsection, we evaluate transmission performance with the proposed monitor. In this setup, a FullTr transmission is performed at the beginning, and subsequent transmissions are triggered according to the monitor’s decision. We first analyze overall performance trends, then provide a detailed example from the basic scenario, including transmission times, bandwidth cost, and average metrics. Finally, typical frames are examined to highlight the advantages and limitations of the proposed monitor.  

Fig.~\ref{SecB_per}(a) shows the LPIPS performance of competing methods under different channel conditions. When the SNR is 5 dB and only PartTr is enabled by the monitor [denoted as SNR = 5 dB (PartTr)], PartTr transmissions are triggered at timeslots 6, 9, 11, and 16. When the SNR drops to 0 dB, more PartTr transmissions are required, which demonstrates the adaptiveness of the proposed monitor. However, when FullTr is controlled by the monitor, performance is also good, but too many transmissions are triggered after the 8th timeslot. This result indicates an incompatibility between the monitor and FullTr.  

Fig.~\ref{SecB_per}(b) evaluates depth-based performance and reveals potential limitations of the proposed monitor. Since depth maps capture only partial image semantics and are affected by transmission errors, depth performance is unstable even when the monitor attempts to maintain errors below the threshold of 0.3. For example, at the 9th timeslot under SNR = 0 dB (PartTr), the depth error exceeds the threshold, yet the corresponding LPIPS value remains acceptable. Moreover, the consistency between depth and LPIPS performance for FullTr is poor. Transmission errors in FullTr often degrade structural semantics in the image, while PartTr better preserves structure by transmitting segmentation maps.

\begin{table} 
	\centering	
	\footnotesize
	\caption{Performance of different transmission methods with the WFM.}
	
	\begin{tabular}{crrrcc}
		\toprule
		Method& SNR&Times&Bandwidth  & LPIPS $\downarrow$ & $\delta>1.25 \downarrow$  \\\midrule
		\multirow{3}{*}{PartTr }&10 dB&4&\textbf{4KBytes}&0.2388 &\textbf{0.1775}\\ \cmidrule{2-6}
		&5 dB&6& 5KBytes &\textbf{0.2330}&0.1857\\ \cmidrule{2-6}
		&0 dB&10& 7KBytes&0.2330&0.2127\\ \midrule
		FullTr &	5 dB &13& 26KBytes&0.1865&0.4204\\ 
		
		\bottomrule
	\end{tabular}
	\label{Metric1}
\end{table}

The results in Table~\ref{Metric1} highlight the differences between FullTr and PartTr when used with the monitor. The total bandwidth consumption for 20 frames is reported. At higher SNRs, fewer bytes are required because WFM prediction alone is sufficient for most timeslots. With the help of the monitor, the average LPIPS values remain stable across SNRs. However, FullTr exhibits worse LPIPS scores than other methods because its noisy reconstructed frames reduce the accuracy of the depth predictor used by the monitor. As a result, the monitor frequently issues unnecessary transmission commands, leading to excessive bandwidth usage. In contrast, PartTr maintains efficiency by leveraging segmentation masks, which align better with the monitor’s depth-based criterion.

In summary, the proposed monitor effectively adapts transmission frequency under varying SNR conditions. Nevertheless, it is more suitable for PartTr than FullTr due to its reliance on depth prediction. To overcome this limitation and better handle camera mobility, an active transmission strategy is investigated in the next subsection.  

\subsection{Performance With Active Adaption}

In this subsection, the transmission strategy based solely on the proposed monitor is referred to as \textit{passive}. We first compare the performance of active and passive strategies in a scenario where the camera moves between two BSs. Then, visual examples are provided to illustrate the superiority of the active strategy. Finally, results from different evaluation metrics are discussed to highlight the advantages and trade-offs of the proposed method.

\begin{figure}[h]
	\centering
	\subfigure[]{
		
		{\includegraphics[width=0.9\linewidth]{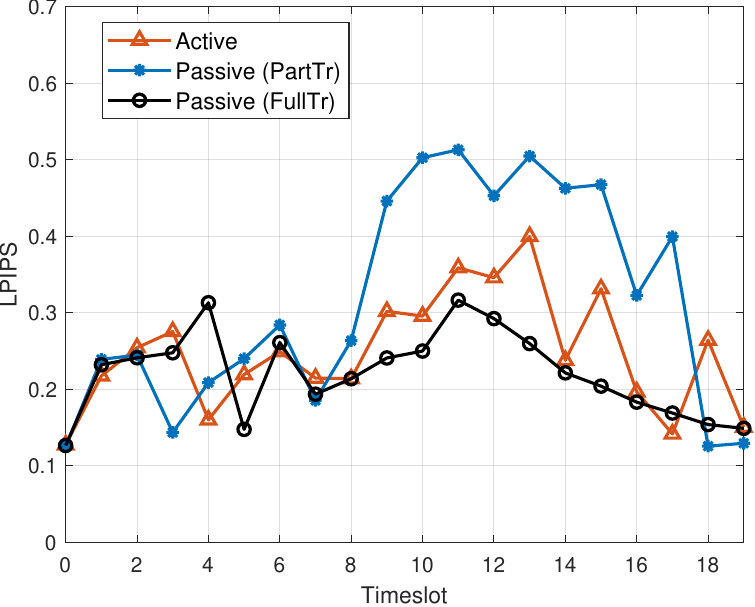}}}\\
	 \subfigure[]{
	 	{\includegraphics[width=0.9\linewidth]{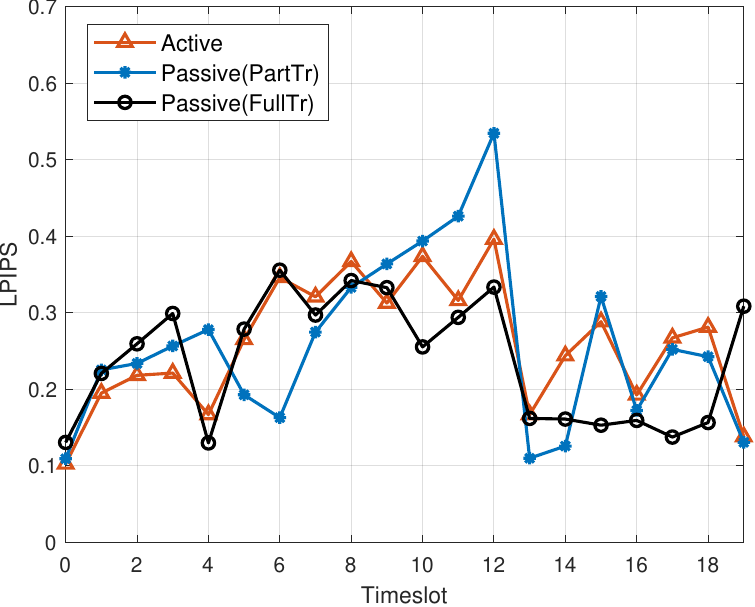}}}
	\caption{Performance of the competing methods with different strategies. (a) From one BS to the other. (b) Sudden interference.}
	\label{SecC_per}
\end{figure}

\begin{figure*}[!h]
	\centering
	
	\subfigure[Passive(FullTr)]{
		\includegraphics[width=0.3\linewidth]{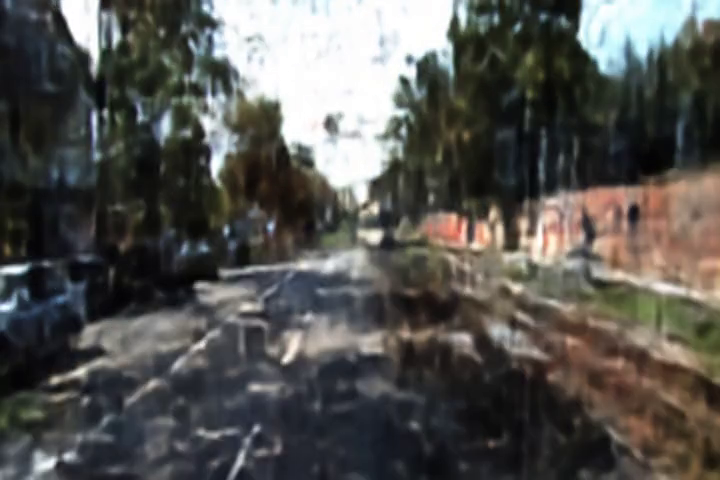}}
	\subfigure[Passive(PartTr)]{
		\includegraphics[width=0.3\linewidth]{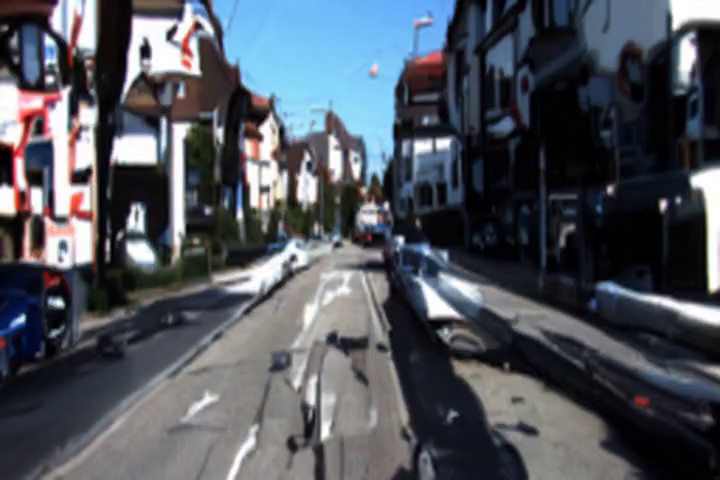}}
	\subfigure[Active]{
		\includegraphics[width=0.3\linewidth]{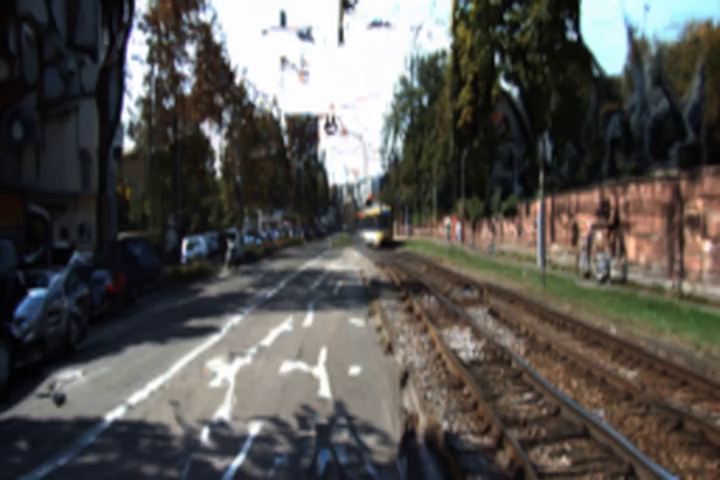}}
	
	\caption{Examples of the WFM-based method using different strategies at timeslot=9.}
	\label{EX_B}
\end{figure*}

Fig.~\ref{SecC_per}(a) shows the performance of competing strategies.  In this experiment, the camera travels along a straight path from $(-100,50)$ to $(100,50)$. The receiver connects to the BS with the stronger SNR, and a handover occurs near the midpoint (timeslot = 10). The transmit power $P_t$ and antenna gain $G$ in (\ref{PG}) are set to 10 and 15, respectively. For \textit{Passive(FullTr)}, the monitor triggers FullTr transmissions. This strategy is overly sensitive to depth changes, resulting in continuous transmissions   around the midpoint where SNR is weak. Thus, although \textit{Passive(FullTr)} achieves the best LPIPS performance here, to show the full picture, transmission overhead values should be also provided. For \textit{Passive(PartTr)}, prediction errors are repaired using PartTr. However, the poor channel condition around the midpoint reduces repair accuracy, and subsequent performance remains poor until the channel improves. By contrast, the proposed \textit{active} strategy schedules two FullTr transmissions at timeslots 7 and 14, achieving performance close to Passive(FullTr) while requiring fewer transmissions. 

Fig.~\ref{SecC_per}(b) shows the channel conditions with a sudden inference, where the LoS path is blocked between timeslots 7 and 13. The active strategy choose timeslots 7, 9 and 11 for FullTr under this sudden condition. Compared to the predictable channel conditions in Fig. \ref{SecC_per}(a), the active method cannot transmit the whole image before the channel becomes poor. In general, the proposed active method still tries to achieve a more stable LPIPS compared to Passive(PartTr). Meanwhile, the proposed active method performs much better than Passive(FullTr) with  similar LPIPS performances and this will be discussed in the following.

\begin{table} 
	\centering	
		\footnotesize
	\caption{Average performances of different WFM-based strategies.}
	
	\begin{tabular}{llrccc}
		\toprule
		 &Method&Bandwidth&SSIM$\uparrow$  &LPIPS$\downarrow$ &Dreamsim $\uparrow$  \\\midrule
	 \multirow{3}{*}{(a)}&Active&10KBytes&0.55&0.248 &\textbf{0.85}\\ \cmidrule{2-6}
		& Passive(PartTr)&\textbf{8KBytes}&0.51 & 0.313&0.77\\ \cmidrule{2-6}
		& Passive(FullTr)&30KBytes&0.75&0.221&0.83\\ \midrule
		 \multirow{3}{*}{(b)}&Active&\textbf{9KBytes}&0.58&0.258
		  &\textbf{0.81}\\ \cmidrule{2-6}
		 &Passive(PartTr)&10KBytes& 0.57&0.257&0.78\\ \cmidrule{2-6}
		 &Passive(FullTr)&28KBytes& 0.73&0.233&\textbf{0.81}\\ 
		
		\bottomrule
	\end{tabular}
	\label{Metric2}
\end{table}

Table~\ref{Metric2} summarizes the performance of different strategies. PartTr, which transmits only structural information, achieves the smallest bandwidth consumption and the best SSIM score, reflecting structural consistency. FullTr, which transmits entire frames, achieves the best LPIPS performance but consumes the most bandwidth. The active strategy provides a balanced compromise: it requires only slightly more bandwidth than Passive(PartTr) while achieving performance close to Passive(FullTr). Importantly, the active strategy achieves the best DreamSim score, indicating superior perceptual and semantic quality. 

Visual examples in Fig.~\ref{EX_B} further highlight the advantages of the active method. Under low SNR, frames received with Passive(FullTr) become blurry due to noise, and subsequent predictions degrade. Passive(PartTr) produces sharper frames using the generative capability of the diffusion model, but transmission noise may alter object semantics. The active method combines the strengths of both approaches: it transmits key frames with FullTr before channel quality deteriorates to preserve accuracy, and subsequently relies on prediction and generation to maintain clarity.  
 
In summary, the proposed active strategy effectively integrates different transmission methods to achieve robust performance in dynamic scenarios. By proactively scheduling transmissions before channel conditions worsen or disconnections occur, the active strategy leverages the long-term predictive capability of WFMs to ensure both reliability and efficiency. The following subsection further explores the potential of this framework through ablation studies.

\subsection{Effectiveness of the Proposed Framework, Ablation Study, and Complexity Analysis}

This subsection first compares the performance of the proposed transmission modules with conventional methods. Then, an ablation study is conducted to demonstrate the effectiveness of introducing WFM. Finally, future research directions are discussed.  

\begin{table} 
	\centering	
	\footnotesize
	\caption{Conventional metrics of the competing methods under varying SNRs.}
	
	\begin{tabular}{llcccc}
		\toprule
		&SNR(dB) &-5&0  & 5 & 10   \\\midrule
		\multirow{4}{*}{MSE}	&AV1+LDPC&0.9867&1.0074&0.0107 &0.0007\\ \cmidrule{2-6}
		&PartTr&0.0908&0.0611&0.0584&0.0644\\ \cmidrule{2-6}
		&FullTr 	 &0.0147& 0.0049&0.0027&0.0020\\ \midrule
		\multirow{4}{*}{LPIPS}	&AV1+LDPC&1.0222&0.9543&0.4388 &0.0471\\ \cmidrule{2-6}
		&PartTr&0.3919& \textbf{0.2458}&\textbf{0.2004}&\textbf{0.2219}\\ \cmidrule{2-6}
		&FullTr 	 &\textbf{0.2856}& 0.1507&0.0993&0.0835\\ 
		
		\bottomrule
	\end{tabular}
	\label{Metric0}
\end{table}

Table~\ref{Metric0} compares the basic performance of the proposed and conventional methods. Conventional approaches fail under low SNRs, where errors exceed the correction capability of channel coding. In such cases, conventional methods often require doubling the bandwidth by lowering modulation order or code rate. By contrast, the proposed framework flexibly combines transmission and generation methods to significantly reduce overhead. For example, PartTr repairs predicted frames with the diffusion model. When SNR $>0$ dB, PartTr achieves good performance. At very low SNR (e.g., $-5$ dB), FullTr remains effective due to its end-to-end design. Overall, the proposed framework adapts transmission strategies to balance performance and bandwidth consumption.

\begin{figure}[!h]
	\centering
	\subfigure[]{
		
		{\includegraphics[width=0.9\linewidth]{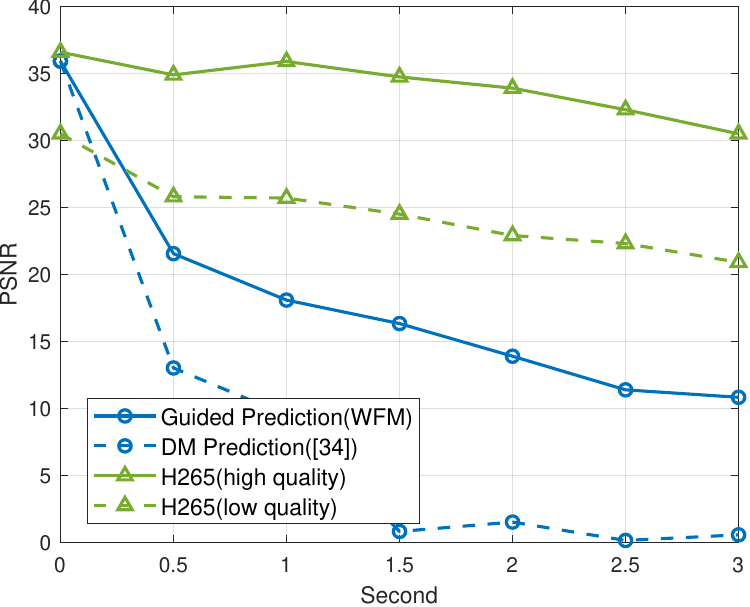}}}\\
	\subfigure[]{
		{\includegraphics[width=0.9\linewidth]{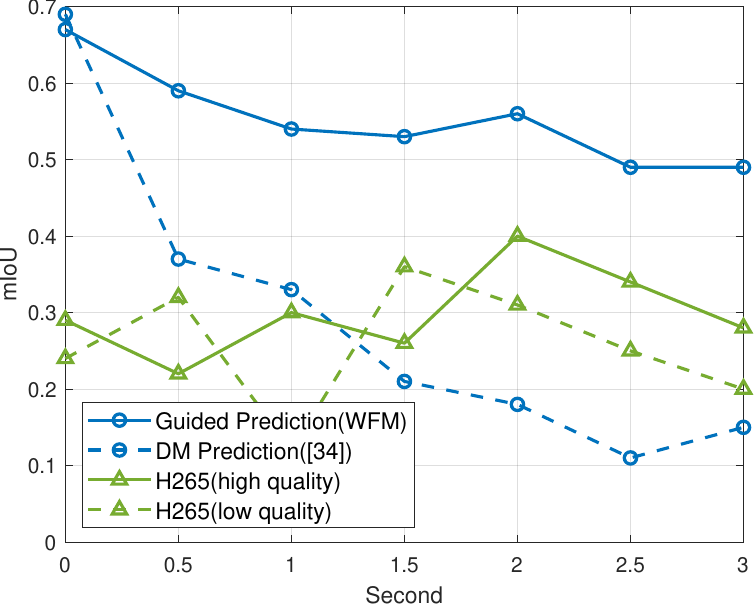}}}
	\caption{Performance of WFM-based and conventional methods. (a) Frame quality over time. (b) mIoU performance.} 
	\label{SecD_aba}
\end{figure}

Fig.~\ref{SecD_aba} presents the ablation study of introducing WFM. Conventional video coding (e.g., H.264/H.265) also predicts subsequent frames and transmits residual information. To compare bandwidth usage, videos are divided into 3-second segments with six frames each. H.264 (high quality) requires 10 KBytes, and H.265 (low quality) requires 5 KBytes. In contrast, FM-based prediction transmits only the first frame and predicts the next five, requiring 2 KBytes. Unlike conventional prediction, the WFM leverages both pretrained world knowledge and text guidance. For comparison, the DM architecture similar to \cite{hoppe2022diffusion} is trained directly for prediction without world knowledge.  

As shown in Fig.~\ref{SecD_aba}(a), conventional video coding degrades frame quality slowly because residuals compensate prediction errors, but it always incurs higher bandwidth. DM-only prediction  degrades more rapidly than WFM-based prediction. This demonstrates that pretrained WFMs exploit physical priors and textual context for more accurate frame generation. Fig.~\ref{SecD_aba}(b) further shows that WFM achieves superior mIoU performance, while conventional coding produces blurrier images that degrade task-level accuracy (e.g., object recognition).  

 \begin{table} 
 	\centering	
 	 \footnotesize
 	 \caption{Complexity of different modules.}
 	
 	\begin{tabular}{lrl}
 		\toprule
 		Module&GPU Memory &Runtime \\\midrule
 		PartTr(Encoder-Decoder)&0.2GB & 1.9 ms \\ \midrule
            PartTr(Segment) & 3GB& 0.8 s\\ \midrule
            PartTr(DM)& 2GB & 3.3 s\\ \midrule
 		FullTr& 0.3GB & 2.3 ms \\ \midrule
            Depth(Detection+Transmission) & 0.8G &0.6 s \\ \midrule
 		WFM' prediction &15GB& 3.6 s\\ 

 		\bottomrule \\
 		\multicolumn{3}{l}{*Testbed is equipped with a CPU (Intel i9-14900k) and}\\
 		\multicolumn{3}{l}{~~a GPU (Nvidia RTX 4090).} 
 
 	\end{tabular}
 	\label{Metric3}
 \end{table}

TABLE \ref{Metric3} presents the complexity metrics of various modules, showing significant differences in GPU memory usage and runtime. For instance, the WFM' prediction module demands a substantial 15G of GPU memory and takes 3.6 seconds to run, while the FullTr module requires only a minuscule 2.3 ms of runtime and negligible GPU memory. The high complexity of some modules can bring a superior in bandwidth cost and video performance. Video transmission primarily focuses on efficiently encoding, transmitting, and decoding video data. However, the presence of alternatives such as FullTr with relatively lower complexity offers a viable solution. By strategically employing less complex modules, we can strike a balance between performance and resource utilization. While the current high complexity issue exists, it is a solvable problem that will be addressed over time with the continuous progress of technology.

Although the pretrained WFM used in this study is constrained by model size and computational resources, recent advances such as Genie3 \cite{WinNT} and Sora2 \cite{liu2024sora} demonstrate that modern WFMs can generate stable predictions over several minutes.   Meanwhile, the memory  and time cost can also be solved by some lightweight design and acceleration techniques, such as Wan2.2 \cite{wan2025} and Cosmos \cite{agarwal2025cosmos}. These rapid developments suggest that WFM-based transmission systems hold strong potential to become highly competitive in the future.

\section{Conclusion}
\label{s6}

This study investigated the use of WFMs to reduce video transmission overhead by leveraging their ability to capture world knowledge. By combining the current frame with textual guidance, WFMs can predict subsequent frames, but the prediction quality is affected by the transmitted content and channel conditions. To address this issue, a depth-based feedback mechanism was proposed to determine whether transmission is required. With the aid of feedback and two optional forward transmission methods, the overall bandwidth consumption of the framework was significantly reduced.  
Furthermore, an active strategy was developed for mobile scenarios by exploiting the trajectory of the camera between BSs. This strategy enables proactive scheduling of transmissions, striking a balance between reconstruction quality and bandwidth cost by flexibly selecting timeslots for transmission or prediction. Finally, the ablation study demonstrated the superiority of introducing WFMs. With the rapid advancement of foundation models, WFM-based communication frameworks are expected to become increasingly competitive and practical in future wireless systems.

	\bibliographystyle{IEEEtran}
	\bibliography{SCWFM}
	
	%
	
	
	
\end{document}